\documentclass[useAMS,usenatbib]{aa}

\usepackage{xcolor}
\usepackage{graphicx}
\usepackage{subcaption}
\usepackage{orcidlink}
\usepackage{float}

\newcommand{\fr}[1]{\begin{color}{red}\ \ifthenelse{\equal{#1}{}} {$\bullet\bullet\bullet$} {$\bullet$\ #1 $\bullet$}\end{color}}

\begin{document}

\title{The multiple origins of spurs in stellar streams}
\subtitle{A challenge for dark matter diagnostics}

\author{Claire~Guillaume\thanks{\email{claire.guillaume@astro.unistra.fr}}\inst{\ref{sxb}}\orcidlink{0009-0005-6540-0229}
\and Alicia~Rivero\inst{\ref{IAC},\ref{Uni_Laguna}}\orcidlink{0009-0001-3808-8074}
\and Titouan~Le Guillois\inst{\ref{sxb}}
\and Florent~Renaud\inst{\ref{sxb},\ref{usias}}\orcidlink{0000-0001-5073-2267}
\and Guillaume~F.~Thomas\inst{\ref{IAC},\ref{Uni_Laguna}}\orcidlink{0000-0002-2468-5521}
\and Nicolas~F.~Martin\inst{\ref{sxb},\ref{MPI}}\orcidlink{0000-0002-1349-202X}
}

\institute{Universit\'e de Strasbourg, CNRS, Observatoire Astronomique de Strasbourg, UMR 7550, 67000 Strasbourg, France\label{sxb}
\and University of Strasbourg Institute for Advanced Study, 5 all\'ee du G\'en\'eral Rouvillois, F-67083 Strasbourg, France\label{usias}
\and Max-Planck-Institut für Astronomie, Königstuhl 17, 69117 Heidelberg, Germany\label{MPI}
\and Instituto de Astrof\'isica de Canarias, E-38205 La Laguna, Tenerife, Spain\label{IAC}
\and Universidad de La Laguna, Dpto. Astrofísica, E-38206 La Laguna, Tenerife, Spain\label{Uni_Laguna}
}

\abstract{Spurs along stellar streams are often interpreted as signatures of interactions with dark matter subhalos, and thus have been proposed as diagnostics on the clustering and nature of dark matter. Although galactic baryonic components could play similar roles as gravitational perturbers of streams, the possible signatures they leave and the degeneracy with subhalos have been overlooked. In this paper, we analyse simulations of streams experiencing disruptions by dark matter subhalos, a galactic bar, and a galaxy merger. Although spurs are more frequently found to be induced by the galaxy merger, the spurs originating from all these processes are morphologically similar. Therefore, such morphological features should not be used alone to characterise the mechanism of their origin, nor to constrain the nature of dark matter.}


\maketitle

\section{Introduction}

Simulations of stellar streams in the $\Lambda$CDM framework have shown that subhalo flybys leave measurable imprints, such as gaps, kinks, and spurs \citep[e.g.,][]{Ibata2002, Johnston2002, Yoon2011, Carlberg2012, Carlberg2013, Erkal2015a, Erkal2015b, Erkal2016, Bonaca2019}.  Inhomogeneities and gaps along streams can also arise from internal dynamics, through epicyclic motions of escaping stars \citep{Kupper2010, Mastrobuono-Battisti2012, Ibata2020}, from baryonic perturbers such as galactic bars, globular clusters, and galactic mergers \citep{Erkal2017, Pearson2017, Ferrone2025, Guillaume2026}, and artificially from inhomogeneous observations \citep{Thomas2016}. Spurs, as features extending off the main stream track, constitute a complementary signature to gaps. Specific spurs in GD-1 and Ophiuchus have been scrutinized in great detail with the goal of associating them to a perturber. In GD-1, the spur has been revealed by Gaia astrometry \citep{PriceWhelan2018} and interpreted as evidence of an impact with a compact dark matter subhalo \citep{Bonaca2019}. Orbital modelling further suggested that the perturber could be a globular cluster or a dark matter subhalo associated with the Sagittarius galaxy \citep{Bonaca2020}. An interaction of GD-1 with Sagittarius has also been considered \citep{deBoer2020, Dillamore2022}, but a Sagittarius or a globular cluster trigger for this spur is now considered unlikely \citep{Bonaca2025}. A spur has also been identified in the Ophiuchus stream \citep{Caldwell2020}, and recently modelled as tidal debris scattered by the decelerating galactic bar \citep{Yang2025}, i.e. without invoking subhalos. 

Current and future large observational programs from the Rubin Observatory \citep{LSST2019}, combined with spectroscopic surveys such as 4MOST \citep{4MOST2019}, DESI \citep{DESI2016}, and WEAVE \citep{WEAVE2024}, will soon detect new streams and map known ones with unprecedented detail. Statistical pipelines are already built to infer the impact of individual dark matter subhalos \citep{Lu2026a, Lu2026b}. However, the other perturbations to the streams' morphologies and possible degeneracies are often overlooked.

In this letter, we use $N$-body simulations of streams evolved under three perturbations: a galactic merger, a galactic bar, and a population of dark matter subhalos, to test whether these conditions trigger spur formation, and to identify possible degeneracies.

\section{Numerical methods}
\label{sec:methods}

We model the same 36 stellar streams as in \citet{Guillaume2026}, using the same globular cluster model, i.e. a \citet{Plummer1911} profile with a particle mass of $6.5\,{\rm M}_\odot$. Each cluster is integrated independently for 5~Gyr. We use \texttt{pyfalcon}, a Python interface to the falcON tree code \citep{Dehnen2002} with a softening length of 1\,pc to compute the self-gravity of the cluster particles, and \texttt{Agama} \citep{Vasiliev2019} to compute the external acceleration from the fitted galactic potentials, as described below. We integrate the motion of particles with a leapfrog scheme with a fixed time step of 50\,kyr. We checked the accuracy of the integration by monitoring energy conservation for streams integrated in a static potential. Spurs are identified by visual inspection of the streams.

To generate the external potentials, we start from the collisionless $N$-body simulations of \citet{Guillaume2026}, modelling a Milky Way-mass galaxy undergoing a galactic merger with a mass ratio of 1:10. The potential of each galaxy is then fitted following \citet{Petersen_Penarrubia2021} and \citet{Lilleengen2023}, with a time-dependent multipole expansion up to the order 10 (Rivero et al., in prep.) using \texttt{Agama}. The fit uses all particles within 400\,kpc of the centre of mass of each galaxy. The total potential is the sum of the two galaxies, fitted separately. We repeat this procedure for the Milky Way-mass galaxy in isolation, in which a bar develops naturally and becomes visible after 2\,Gyr. The bar reaches a radius of $\sim$8~kpc, with a pattern speed decreasing from $27$ to $17~\mathrm{km\,s^{-1}\,kpc^{-1}}$ between $\sim 3$ and 5~Gyr. Additionally, by fitting the isolated galaxy before the bar forms, we obtain a model for an axisymmetric disc, used as a reference below. Finally, we create a fourth model by converting 10\% of the dark matter mass of the isolated galaxy into a population of subhalos. The subhalo mass follows $\mathrm{d}N/\mathrm{d}M \propto M^{-1.9}$ between $10^5$ and $10^9\,{\rm M}_\odot$ \citep{Springel2008}. Each subhalo is represented by a single particle exerting a \citet{Plummer1911} force on the streams in accordance with the \texttt{pyfalcon} force kernel. The Plummer scale radius is set as in \citet{Erkal2016} from the Via Lactea II simulation \citep{Diemand2008} as $1.62\,{\rm kpc}\,(M_{\rm sub}/10^8\,{\rm M}_\odot)^{0.5}$, giving $\simeq 51$\,pc for $10^5\,{\rm M}_\odot$ and $\simeq 5.1$\,kpc for $10^9\,{\rm M}_\odot$ subhalos. The initial positions and velocities of the subhalos are drawn randomly from the initial dark matter particles within 400\,kpc in the $N$-body simulation. Each subhalo is first integrated alone in the fitted potential, and its trajectory is stored. We then integrate each stellar stream in the same fitted potential including the subhalos along their precomputed trajectories. The gravitational effect of the streams on subhalos is thus neglected. Using Plummer rather than NFW subhalos \citep{NFW1997} may underestimate the number of spurs, as NFW subhalos produce twice as many gaps \citep{Menker2026}, although fewer than 1\% of encounters probe the density profile. Conversely, neglecting structural evolution, tidal mass loss and dynamical friction \citep{Errani2020} possibly overestimates the subhalo impact.

\section{Results}
\label{sec:results}

\begin{figure}
  \centering
  \includegraphics[width=\columnwidth]{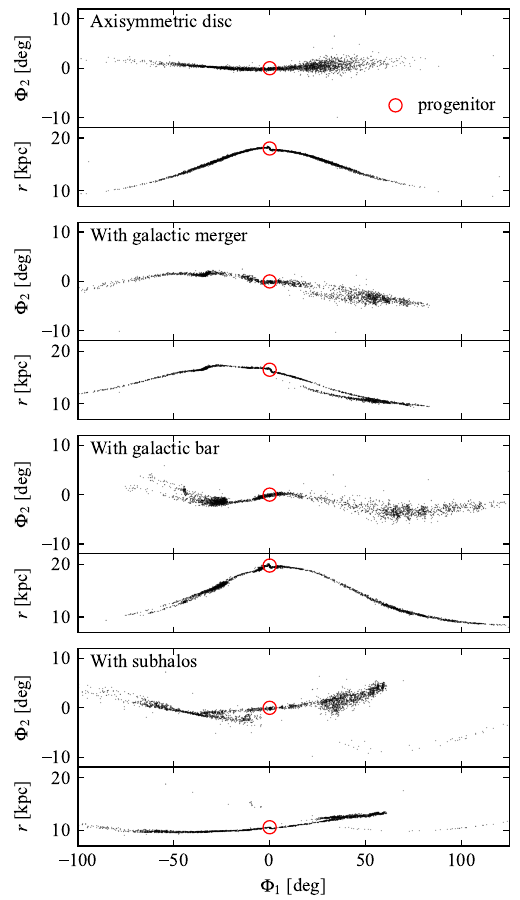}
  \caption{The same stream shown in four configurations (top to bottom): axisymmetric disc, galactic merger, galactic bar, and with dark matter subhalos, in great-circle coordinates, with $\Phi_2$ (top) and galactocentric distance $r$ (bottom) shown as a function of $\Phi_1$.}
  \label{fig:example}
\end{figure}

\begin{figure}
  \centering
  \includegraphics[width=\columnwidth]{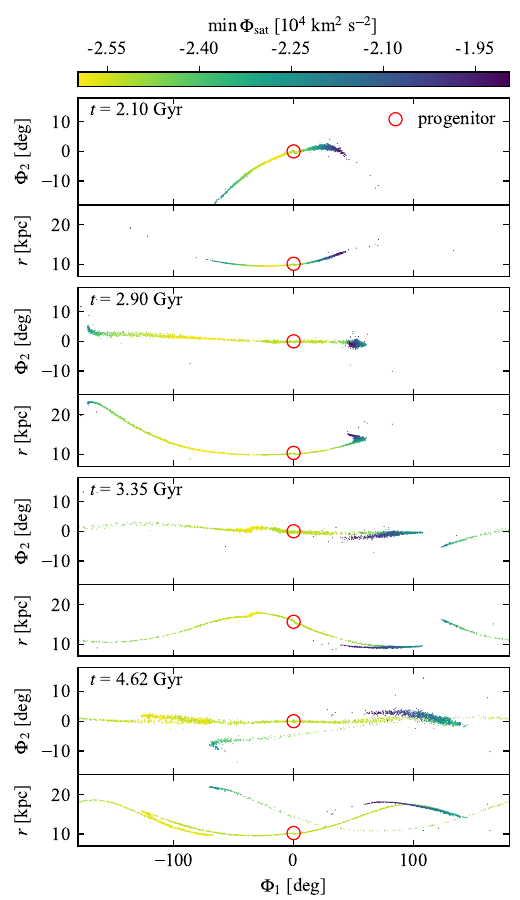}
  \caption{Evolution of the stream from Fig.~\ref{fig:example} in the galactic merger scenario, at four epochs. Particles are coloured by the minimum values of the satellite potential they experienced during the course of the merger.}
  \label{fig:times}
\end{figure}

\begin{figure*}
  \centering
  \includegraphics[width=\textwidth]{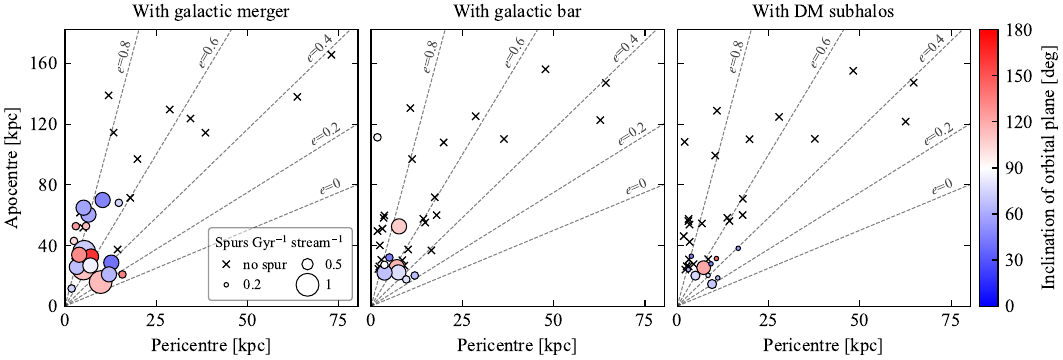}
  \caption{Pericentre and apocentre of the 36 stellar streams in the three scenarios: galactic merger, barred galaxy, and with dark matter subhalos. Black crosses mark the streams that do not form any visible spur. Coloured circles mark the ones that form at least one spur: the colour gives the inclination of the orbital plane, and the size gives the spur formation rate per stream per Gyr. Dashed lines indicate constant eccentricities.}
  \label{fig:apopericentre}
\end{figure*}

\subsection{Spurs created by a galactic merger}
\label{sec:merger}

The first and second rows of Fig.~\ref{fig:example} show the same stream in the axisymmetric, unperturbed disc and in the merger configuration respectively. For completeness, Fig.~\ref{fig:A3} shows its radial and proper motions. A spur is visible at the leading end ($\Phi_1 > 0$), while being absent in the axisymmetric case. Two gaps are also present, at $\Phi_1 \sim -20^\circ$ and $25^\circ$. This combination of a spur and gaps resembles the morphology of the GD-1 stream \citep{PriceWhelan2018, Bonaca2019}. Therefore, our study adds mergers to the list of possible origins of such observed features. However, a deeper exploration of the parameter space is needed before concluding on the uniqueness of this scenario to explain the observed structure of GD-1.

To understand how this spur forms, we show this stream at four epochs in Fig.~\ref{fig:times}. The first panel displays the moment of first pericentre passage of the satellite galaxy, which also corresponds to the closest encounter of the stream and the satellite. The stars at the leading end experience a shallower potential from the satellite than the rest of the stream. Shortly after the second pericentre passage ($2.90$\,Gyr), these stars are offset from the main stream, forming a spur at $\Phi_1 \sim 50^\circ$, best visible in the $\Phi_1$--$r$ projection. With time, the spur elongates (third and fourth epochs). A second spur appears near $\Phi_1 \sim -30^\circ$ at $3.35$\,Gyr and near $\Phi_1 \sim -120^\circ$ at $4.62$\,Gyr, made of stars having experienced a deep satellite potential. Both spurs consist of stars whose interaction with the satellite strongly differs from that of the rest of the stream. This differential gravitational force from the satellite along the stream induces the kinetic and orbital deviations observed as spurs.

Streams on smaller orbits undergo multiple interactions with the satellite and the main galactic disc, yielding strong perturbations and more complex morphologies than the illustration of Fig.~\ref{fig:example}. Half of the streams in the merger simulation develop one or several visible spurs, as shown in Fig.~\ref{fig:apopericentre}. Streams on wide orbits are less likely to be significantly perturbed by the infalling satellite and galactic disc and thus yield no spurs.

\subsection{Spurs created by the galactic bar}
\label{sec:bar}

Through differential gravitational torques, a galactic bar can also perturb stellar streams and produce gaps and asymmetric tidal tails \citep[e.g.,][]{Erkal2017, Pearson2017}. Here, we show that the bar-induced spur proposed for Ophiuchus \citep{Yang2025} is not unique to that stream but a generic outcome of bar--stream interactions. The third row of Fig.~\ref{fig:example} shows the same stream as before but now evolved in the barred galaxy potential, where it develops a spur in its trailing arm at $\Phi_1 \sim -50^\circ$. 

Of the 36 streams in the bar setup, 25\% form a spur, and all after the bar has formed. As shown in Fig.~\ref{fig:apopericentre}, these have pericentres smaller than 11\,kpc, and are preferentially on prograde orbits, with a lower formation rate than in the merger case. However, among all the 24 streams with a pericentre smaller than 11\,kpc, only 9 develop a spur. Therefore, while the presence of a bar is a necessary condition, a passage near it is not sufficient to trigger spur formation. Other factors such as the orbital phase of the stream relative to the bar and the inclination of the orbital plane must be taken into account. These aspects will be explored in a forthcoming dedicated study of the parameter space.

\subsection{Spurs created by dark matter subhalos}
\label{sec:dm}

The last row of Fig.~\ref{fig:example} shows that our stream example yields two spurs ($\Phi_1 \approx -25^\circ$ and $25^\circ$) in the configuration with dark matter subhalos. Among the 36 stellar streams, 28\% develop at least one visible spur in this simulation, all of them with apocentres shorter than 40\,kpc (Fig.~\ref{fig:apopericentre}). The spatial distribution of subhalos follows the dark matter profile of the galaxy. They are thus more numerous toward the centre. Therefore, a stream with a small apocentre and thus a short orbital period experiences numerous and diverse encounters with subhalos and is more likely to yield spurs than more distant streams. Yet, the streams in this case have the lowest spur formation rate per stream.

Here, the nature of the perturbation differs from the merger case. The satellite galaxy is a single, massive and spatially extended perturber that acts on a large portion of a given stream at once. The tidal effect thus covers larger scales. By contrast, encounters with the more numerous but smaller and less massive dark matter subhalos disturb only short segments of the stream, on local, smaller scales. A stream can therefore accumulate the distinct signatures of several impacts at different positions and times, caused by subhalos with a wide range of masses and impact parameters. Consequently, while the encounter with a single subhalo is sufficient to trigger the formation of a spur \citep{Bonaca2019}, the repetition and diversity of encounters in a galactic halo makes it impossible to associate a given spur with a particular event or subhalo, or to determine if more than one subhalo has been involved in this formation.

\section{Discussion and conclusion}
\label{sec:discussion}

Using $N$-body simulations of a population of stellar streams, we show that spurs are not a signature unique to interactions with dark matter subhalos. A galaxy merger, a galactic bar, and a population of dark matter subhalos all produce spurs, in respectively 50, 25, and 28\% of the streams modelled. Our simulations demonstrate that each process can produce morphologically similar spurs. The high fraction of streams showing a spur, and the high formation rate caused by a galaxy merger suggest that observed spurs on the streams around the Milky Way may have been caused by recent or ongoing interactions with Sagittarius and/or the Large Magellanic Cloud.

Although spiral arms and giant molecular clouds are not included in our simulations, they could also produce spurs through the same mechanism of tidal perturbation. This degeneracy implies that the presence of a spur in an observed stream should not be taken as evidence for dark matter subhalos, unless all other types of perturbers have been ruled out. Disentangling the signatures of different perturbers is challenging for streams with small pericentres, since these are simultaneously exposed to strong effects from the Milky Way itself (bar, spiral arms, giant molecular clouds), and have a higher probability of having interacted with infalling satellites and recent mergers. 

A statistical approach applied to the broader stream population may nonetheless offer a viable path forward: rather than attributing individual features to individual perturbers, an aggregate measure of disruption across many streams, compared to predictions from simulations, could still constrain the abundance and properties of dark matter subhalos. Such a measure should combine morphology with kinematics as different dark matter models may imprint both properties differently \citep{Carlberg2024}. Suites of simulations spanning a wider range of subhalo populations, bar strengths, and merger histories, combined with a systematic characterisation of the resulting stream properties, would be needed to build a holistic theoretical framework.

\begin{acknowledgements}
Co-funded by the European Union (Widening Participation, ExGal-Twin, GA 101158446). GFT acknowledges support from the Agencia Estatal de Investigaci\'on del Ministerio de Ciencia e Innovaci\'on (AEI-MCIN) under grant number RYC2024-051016-I funded by MCIN/AEI/10.13039/501100011033 and by the European Social Fund Plus. GFT and AR acknowledge support from the Agencia Estatal de Investigaci\'on del Ministerio de Ciencia e Innovaci\'on (AEI-MCIN) under grant number PID2023-150319NB-C21.
\end{acknowledgements}

\bibliographystyle{aa}
\bibliography{biblio}

@ARTICLE{Guillaume2026,
       author = {{Guillaume}, Claire and {Renaud}, Florent and {Martin}, Nicolas F. and {Famaey}, Benoit and {Di Matteo}, Paola and {Thomas}, Guillaume F. and {Ferrone}, Salvatore and {Ibata}, Rodrigo and {Pagnini}, Giulia},
        title = "{Asymmetries in stellar streams induced by a galactic merger}",
      journal = {\aap},
         year = 2026,
        month = jan,
       volume = {705},
          eid = {A6},
        pages = {A6},
          doi = {10.1051/0004-6361/202557552},
archivePrefix = {arXiv},
       eprint = {2510.06329},
 primaryClass = {astro-ph.GA},
       adsurl = {https://ui.adsabs.harvard.edu/abs/2026A&A...705A...6G}
}

@ARTICLE{Vasiliev2019,
       author = {{Vasiliev}, Eugene},
        title = "{AGAMA: action-based galaxy modelling architecture}",
      journal = {\mnras},
         year = 2019,
        month = jan,
       volume = {482},
       number = {2},
        pages = {1525-1544},
          doi = {10.1093/mnras/sty2672},
archivePrefix = {arXiv},
       eprint = {1802.08239},
 primaryClass = {astro-ph.GA},
       adsurl = {https://ui.adsabs.harvard.edu/abs/2019MNRAS.482.1525V}
}

@ARTICLE{Ibata2020,
       author = {{Ibata}, Rodrigo and {Thomas}, Guillaume and {Famaey}, Benoit and {Malhan}, Khyati and {Martin}, Nicolas and {Monari}, Giacomo},
        title = "{Detection of Strong Epicyclic Density Spikes in the GD-1 Stellar Stream: An Absence of Evidence for the Influence of Dark Matter Subhalos?}",
      journal = {\apj},
         year = 2020,
        month = mar,
       volume = {891},
       number = {2},
          eid = {161},
        pages = {161},
          doi = {10.3847/1538-4357/ab7303},
archivePrefix = {arXiv},
       eprint = {2002.01488},
 primaryClass = {astro-ph.GA},
       adsurl = {https://ui.adsabs.harvard.edu/abs/2020ApJ...891..161I}
}

@ARTICLE{Johnston2002,
       author = {{Johnston}, Kathryn V. and {Spergel}, David N. and {Haydn}, Christian},
        title = "{How Lumpy Is the Milky Way's Dark Matter Halo?}",
      journal = {\apj},
         year = 2002,
        month = may,
       volume = {570},
       number = {2},
        pages = {656-664},
          doi = {10.1086/339791},
archivePrefix = {arXiv},
       eprint = {astro-ph/0111196},
 primaryClass = {astro-ph},
       adsurl = {https://ui.adsabs.harvard.edu/abs/2002ApJ...570..656J}
}

@ARTICLE{Ibata2002,
       author = {{Ibata}, R.~A. and {Lewis}, G.~F. and {Irwin}, M.~J. and {Quinn}, T.},
        title = "{Uncovering cold dark matter halo substructure with tidal streams}",
      journal = {\mnras},
         year = 2002,
        month = jun,
       volume = {332},
       number = {4},
        pages = {915-920},
          doi = {10.1046/j.1365-8711.2002.05358.x},
archivePrefix = {arXiv},
       eprint = {astro-ph/0110690},
 primaryClass = {astro-ph},
       adsurl = {https://ui.adsabs.harvard.edu/abs/2002MNRAS.332..915I}
}

@ARTICLE{Erkal2015a,
       author = {{Erkal}, Denis and {Belokurov}, Vasily},
        title = "{Forensics of subhalo-stream encounters: the three phases of gap growth}",
      journal = {\mnras},
         year = 2015,
        month = jun,
       volume = {450},
       number = {1},
        pages = {1136-1149},
          doi = {10.1093/mnras/stv655},
archivePrefix = {arXiv},
       eprint = {1412.6035},
 primaryClass = {astro-ph.GA},
       adsurl = {https://ui.adsabs.harvard.edu/abs/2015MNRAS.450.1136E}
}

@ARTICLE{Erkal2015b,
       author = {{Erkal}, Denis and {Belokurov}, Vasily},
        title = "{Properties of dark subhaloes from gaps in tidal streams}",
      journal = {\mnras},
         year = 2015,
        month = dec,
       volume = {454},
       number = {4},
        pages = {3542-3558},
          doi = {10.1093/mnras/stv2122},
archivePrefix = {arXiv},
       eprint = {1507.05625},
 primaryClass = {astro-ph.GA},
       adsurl = {https://ui.adsabs.harvard.edu/abs/2015MNRAS.454.3542E}
}

@ARTICLE{Erkal2016,
       author = {{Erkal}, Denis and {Belokurov}, Vasily and {Bovy}, Jo and {Sanders}, Jason L.},
        title = "{The number and size of subhalo-induced gaps in stellar streams}",
      journal = {\mnras},
         year = 2016,
        month = nov,
       volume = {463},
       number = {1},
        pages = {102-119},
          doi = {10.1093/mnras/stw1957},
archivePrefix = {arXiv},
       eprint = {1606.04946},
 primaryClass = {astro-ph.GA},
       adsurl = {https://ui.adsabs.harvard.edu/abs/2016MNRAS.463..102E}
}

@ARTICLE{Erkal2017,
       author = {{Erkal}, Denis and {Koposov}, Sergey E. and {Belokurov}, Vasily},
        title = "{A sharper view of Pal 5's tails: discovery of stream perturbations with a novel non-parametric technique}",
      journal = {\mnras},
         year = 2017,
        month = sep,
       volume = {470},
       number = {1},
        pages = {60-84},
          doi = {10.1093/mnras/stx1208},
archivePrefix = {arXiv},
       eprint = {1609.01282},
 primaryClass = {astro-ph.GA},
       adsurl = {https://ui.adsabs.harvard.edu/abs/2017MNRAS.470...60E}
}

@ARTICLE{Ferrone2025,
       author = {{Ferrone}, Salvatore and {Montuori}, Marco and {Di Matteo}, Paola and {Mastrobuono-Battisti}, Alessandra and {Ibata}, Rodrigo and {Bianchini}, Paolo and {Khoperskov}, Sergey and {Leclerc}, Nicolas and {Hottier}, Clement and {Stein}, Eliot and {Valls-Gabaud}, David and {Owain Snaith}, N. and {Haywood}, Misha},
        title = "{Gaps in stellar streams as a result of globular cluster flybys: The case of Palomar 5}",
      journal = {\aap},
         year = 2025,
        month = jul,
       volume = {699},
          eid = {A289},
        pages = {A289},
          doi = {10.1051/0004-6361/202553923},
archivePrefix = {arXiv},
       eprint = {2502.03941},
 primaryClass = {astro-ph.GA},
       adsurl = {https://ui.adsabs.harvard.edu/abs/2025A&A...699A.289F}
}

@ARTICLE{Dehnen2002,
       author = {{Dehnen}, Walter},
        title = "{A Hierarchical <E10>O</E10>(N) Force Calculation Algorithm}",
      journal = {Journal of Computational Physics},
         year = 2002,
        month = jun,
       volume = {179},
       number = {1},
        pages = {27-42},
          doi = {10.1006/jcph.2002.7026},
archivePrefix = {arXiv},
       eprint = {astro-ph/0202512},
 primaryClass = {astro-ph},
       adsurl = {https://ui.adsabs.harvard.edu/abs/2002JCoPh.179...27D}
}

@ARTICLE{Plummer1911,
       author = {{Plummer}, H.~C.},
        title = "{On the problem of distribution in globular star clusters}",
      journal = {\mnras},
         year = 1911,
        month = mar,
       volume = {71},
        pages = {460-470},
          doi = {10.1093/mnras/71.5.460},
       adsurl = {https://ui.adsabs.harvard.edu/abs/1911MNRAS..71..460P}
}

@ARTICLE{NFW1997,
       author = {{Navarro}, Julio F. and {Frenk}, Carlos S. and {White}, Simon D.~M.},
        title = "{A Universal Density Profile from Hierarchical Clustering}",
      journal = {\apj},
         year = 1997,
        month = dec,
       volume = {490},
       number = {2},
        pages = {493-508},
          doi = {10.1086/304888},
archivePrefix = {arXiv},
       eprint = {astro-ph/9611107},
 primaryClass = {astro-ph},
       adsurl = {https://ui.adsabs.harvard.edu/abs/1997ApJ...490..493N}
}

@ARTICLE{Yoon2011,
       author = {{Yoon}, Joo Heon and {Johnston}, Kathryn V. and {Hogg}, David W.},
        title = "{Clumpy Streams from Clumpy Halos: Detecting Missing Satellites with Cold Stellar Structures}",
      journal = {\apj},
         year = 2011,
        month = apr,
       volume = {731},
       number = {1},
          eid = {58},
        pages = {58},
          doi = {10.1088/0004-637X/731/1/58},
archivePrefix = {arXiv},
       eprint = {1012.2884},
 primaryClass = {astro-ph.GA},
       adsurl = {https://ui.adsabs.harvard.edu/abs/2011ApJ...731...58Y}
}

@ARTICLE{Carlberg2012,
       author = {{Carlberg}, R.~G.},
        title = "{Dark Matter Sub-halo Counts via Star Stream Crossings}",
      journal = {\apj},
         year = 2012,
        month = mar,
       volume = {748},
       number = {1},
          eid = {20},
        pages = {20},
          doi = {10.1088/0004-637X/748/1/20},
archivePrefix = {arXiv},
       eprint = {1109.6022},
 primaryClass = {astro-ph.CO},
       adsurl = {https://ui.adsabs.harvard.edu/abs/2012ApJ...748...20C}
}

@ARTICLE{Carlberg2013,
       author = {{Carlberg}, R.~G.},
        title = "{The Dynamics of Star Stream Gaps}",
      journal = {\apj},
         year = 2013,
        month = oct,
       volume = {775},
       number = {2},
          eid = {90},
        pages = {90},
          doi = {10.1088/0004-637X/775/2/90},
archivePrefix = {arXiv},
       eprint = {1307.1929},
 primaryClass = {astro-ph.GA},
       adsurl = {https://ui.adsabs.harvard.edu/abs/2013ApJ...775...90C}
}

@ARTICLE{Pearson2017,
       author = {{Pearson}, Sarah and {Price-Whelan}, Adrian M. and {Johnston}, Kathryn V.},
        title = "{Gaps and length asymmetry in the stellar stream Palomar 5 as effects of Galactic bar rotation}",
      journal = {Nature Astronomy},
         year = 2017,
        month = aug,
       volume = {1},
        pages = {633-639},
          doi = {10.1038/s41550-017-0220-3},
archivePrefix = {arXiv},
       eprint = {1703.04627},
 primaryClass = {astro-ph.GA},
       adsurl = {https://ui.adsabs.harvard.edu/abs/2017NatAs...1..633P}
}

@ARTICLE{Thomas2016,
       author = {{Thomas}, Guillaume F. and {Ibata}, R. and {Famaey}, B. and {Martin}, N.~F. and {Lewis}, G.~F.},
        title = "{Exploring the reality of density substructures in the Palomar 5 stellar stream}",
      journal = {\mnras},
         year = 2016,
        month = aug,
       volume = {460},
       number = {3},
        pages = {2711-2719},
          doi = {10.1093/mnras/stw1189},
archivePrefix = {arXiv},
       eprint = {1605.05520},
 primaryClass = {astro-ph.GA},
       adsurl = {https://ui.adsabs.harvard.edu/abs/2016MNRAS.460.2711T}
}

@ARTICLE{Kupper2010,
       author = {{K{\"u}pper}, Andreas H.~W. and {Kroupa}, Pavel and {Baumgardt}, Holger and {Heggie}, Douglas C.},
        title = "{Tidal tails of star clusters}",
      journal = {\mnras},
         year = 2010,
        month = jan,
       volume = {401},
       number = {1},
        pages = {105-120},
          doi = {10.1111/j.1365-2966.2009.15690.x},
archivePrefix = {arXiv},
       eprint = {0909.2619},
 primaryClass = {astro-ph.SR},
       adsurl = {https://ui.adsabs.harvard.edu/abs/2010MNRAS.401..105K}
}

@ARTICLE{Dillamore2022,
       author = {{Dillamore}, Adam M. and {Belokurov}, Vasily and {Evans}, N. Wyn and {Price-Whelan}, Adrian M.},
        title = "{The impact of a massive Sagittarius dSph on GD-1-like streams}",
      journal = {\mnras},
         year = 2022,
        month = oct,
       volume = {516},
       number = {2},
        pages = {1685-1703},
          doi = {10.1093/mnras/stac2311},
archivePrefix = {arXiv},
       eprint = {2205.13547},
 primaryClass = {astro-ph.GA},
       adsurl = {https://ui.adsabs.harvard.edu/abs/2022MNRAS.516.1685D}
}

@ARTICLE{Bonaca2019,
       author = {{Bonaca}, Ana and {Hogg}, David W. and {Price-Whelan}, Adrian M. and {Conroy}, Charlie},
        title = "{The Spur and the Gap in GD-1: Dynamical Evidence for a Dark Substructure in the Milky Way Halo}",
      journal = {\apj},
         year = 2019,
        month = jul,
       volume = {880},
       number = {1},
          eid = {38},
        pages = {38},
          doi = {10.3847/1538-4357/ab2873},
archivePrefix = {arXiv},
       eprint = {1811.03631},
 primaryClass = {astro-ph.GA},
       adsurl = {https://ui.adsabs.harvard.edu/abs/2019ApJ...880...38B}
}

@ARTICLE{Bonaca2020,
       author = {{Bonaca}, Ana and {Conroy}, Charlie and {Hogg}, David W. and {Cargile}, Phillip A. and {Caldwell}, Nelson and {Naidu}, Rohan P. and {Price-Whelan}, Adrian M. and {Speagle}, Joshua S. and {Johnson}, Benjamin D.},
        title = "{High-resolution Spectroscopy of the GD-1 Stellar Stream Localizes the Perturber near the Orbital Plane of Sagittarius}",
      journal = {\apjl},
         year = 2020,
        month = apr,
       volume = {892},
       number = {2},
          eid = {L37},
        pages = {L37},
          doi = {10.3847/2041-8213/ab800c},
archivePrefix = {arXiv},
       eprint = {2001.07215},
 primaryClass = {astro-ph.GA},
       adsurl = {https://ui.adsabs.harvard.edu/abs/2020ApJ...892L..37B}
}

@ARTICLE{Bonaca2025,
       author = {{Bonaca}, Ana and {Price-Whelan}, Adrian M.},
        title = "{Stellar streams in the Gaia era}",
      journal = {\nar},
         year = 2025,
        month = jun,
       volume = {100},
          eid = {101713},
        pages = {101713},
          doi = {10.1016/j.newar.2024.101713},
archivePrefix = {arXiv},
       eprint = {2405.19410},
 primaryClass = {astro-ph.GA},
       adsurl = {https://ui.adsabs.harvard.edu/abs/2025NewAR.10001713B}
}

@ARTICLE{Mastrobuono-Battisti2012,
       author = {{Mastrobuono-Battisti}, A. and {Di Matteo}, P. and {Montuori}, M. and {Haywood}, M.},
        title = "{Clumpy streams in a smooth dark halo: the case of Palomar 5}",
      journal = {\aap},
         year = 2012,
        month = oct,
       volume = {546},
          eid = {L7},
        pages = {L7},
          doi = {10.1051/0004-6361/201219563},
archivePrefix = {arXiv},
       eprint = {1209.0466},
 primaryClass = {astro-ph.GA},
       adsurl = {https://ui.adsabs.harvard.edu/abs/2012A&A...546L...7M}
}

@ARTICLE{Springel2008,
       author = {{Springel}, V. and {Wang}, J. and {Vogelsberger}, M. and {Ludlow}, A. and {Jenkins}, A. and {Helmi}, A. and {Navarro}, J.~F. and {Frenk}, C.~S. and {White}, S.~D.~M.},
        title = "{The Aquarius Project: the subhaloes of galactic haloes}",
      journal = {\mnras},
         year = 2008,
        month = dec,
       volume = {391},
       number = {4},
        pages = {1685-1711},
          doi = {10.1111/j.1365-2966.2008.14066.x},
archivePrefix = {arXiv},
       eprint = {0809.0898},
 primaryClass = {astro-ph},
       adsurl = {https://ui.adsabs.harvard.edu/abs/2008MNRAS.391.1685S}
}

@ARTICLE{WEAVE2024,
       author = {{Jin}, Shoko and {Trager}, Scott C. and {Dalton}, Gavin B. and {Aguerri}, J. Alfonso L. and {Drew}, J.~E. and {Falc{\'o}n-Barroso}, Jes{\'u}s and {G{\"a}nsicke}, Boris T. and {Hill}, Vanessa and {Iovino}, Angela and {Pieri}, Matthew M. and et al.},
        title = "{The wide-field, multiplexed, spectroscopic facility WEAVE: Survey design, overview, and simulated implementation}",
      journal = {\mnras},
         year = 2024,
        month = may,
       volume = {530},
       number = {3},
        pages = {2688-2730},
          doi = {10.1093/mnras/stad557},
archivePrefix = {arXiv},
       eprint = {2212.03981},
 primaryClass = {astro-ph.IM},
       adsurl = {https://ui.adsabs.harvard.edu/abs/2024MNRAS.530.2688J}
}

@ARTICLE{4MOST2019,
       author = {{de Jong}, R.~S. and {Agertz}, O. and {Berbel}, A.~A. and {Aird}, J. and {Alexander}, D.~A. and {Amarsi}, A. and {Anders}, F. and {Andrae}, R. and {Ansarinejad}, B. and {Ansorge}, W. and et al.},
        title = "{4MOST: Project overview and information for the First Call for Proposals}",
      journal = {The Messenger},
         year = 2019,
        month = mar,
       volume = {175},
        pages = {3-11},
          doi = {10.18727/0722-6691/5117},
archivePrefix = {arXiv},
       eprint = {1903.02464},
 primaryClass = {astro-ph.IM},
       adsurl = {https://ui.adsabs.harvard.edu/abs/2019Msngr.175....3D}
}

@ARTICLE{Menker2026,
       author = {{Menker}, Paul and {Benson}, Andrew},
        title = "{Advancing stellar streams as a dark matter probe ─ I: effects of subhalo density profile}",
      journal = {\mnras},
         year = 2026,
        month = may,
       volume = {548},
       number = {1},
          eid = {stag538},
        pages = {stag538},
          doi = {10.1093/mnras/stag538},
archivePrefix = {arXiv},
       eprint = {2406.11989},
 primaryClass = {astro-ph.GA},
       adsurl = {https://ui.adsabs.harvard.edu/abs/2026MNRAS.548ag538M}
}

@ARTICLE{Errani2020,
       author = {{Errani}, Rapha{\"e}l and {Pe{\~n}arrubia}, Jorge},
        title = "{Can tides disrupt cold dark matter subhaloes?}",
      journal = {\mnras},
         year = 2020,
        month = feb,
       volume = {491},
       number = {4},
        pages = {4591-4601},
          doi = {10.1093/mnras/stz3349},
archivePrefix = {arXiv},
       eprint = {1906.01642},
 primaryClass = {astro-ph.GA},
       adsurl = {https://ui.adsabs.harvard.edu/abs/2020MNRAS.491.4591E}
}

@ARTICLE{PriceWhelan2018,
       author = {{Price-Whelan}, Adrian M. and {Bonaca}, Ana},
        title = "{Off the Beaten Path: Gaia Reveals GD-1 Stars outside of the Main Stream}",
      journal = {\apjl},
         year = 2018,
        month = aug,
       volume = {863},
       number = {2},
          eid = {L20},
        pages = {L20},
          doi = {10.3847/2041-8213/aad7b5},
archivePrefix = {arXiv},
       eprint = {1805.00425},
 primaryClass = {astro-ph.GA},
       adsurl = {https://ui.adsabs.harvard.edu/abs/2018ApJ...863L..20P}
}

@ARTICLE{deBoer2020,
       author = {{de Boer}, T.~J.~L. and {Erkal}, D. and {Gieles}, M.},
        title = "{A closer look at the spur, blob, wiggle, and gaps in GD-1}",
      journal = {\mnras},
         year = 2020,
        month = jun,
       volume = {494},
       number = {4},
        pages = {5315-5332},
          doi = {10.1093/mnras/staa917},
archivePrefix = {arXiv},
       eprint = {1911.05745},
 primaryClass = {astro-ph.GA},
       adsurl = {https://ui.adsabs.harvard.edu/abs/2020MNRAS.494.5315D}
}

@ARTICLE{Petersen_Penarrubia2021,
       author = {{Petersen}, Michael S. and {Pe{\~n}arrubia}, Jorge},
        title = "{Detection of the Milky Way reflex motion due to the Large Magellanic Cloud infall}",
      journal = {Nature Astronomy},
         year = 2021,
        month = jan,
       volume = {5},
        pages = {251-255},
          doi = {10.1038/s41550-020-01254-3},
archivePrefix = {arXiv},
       eprint = {2011.10581},
 primaryClass = {astro-ph.GA},
       adsurl = {https://ui.adsabs.harvard.edu/abs/2021NatAs...5..251P}
}

@ARTICLE{Lilleengen2023,
       author = {{Lilleengen}, Sophia and {Petersen}, Michael S. and {Erkal}, Denis and {Pe{\~n}arrubia}, Jorge and {Koposov}, Sergey E. and {Li}, Ting S. and {Cullinane}, Lara R. and {Ji}, Alexander P. and {Kuehn}, Kyler and {Lewis}, Geraint F. and {Mackey}, Dougal and {Pace}, Andrew B. and {Shipp}, Nora and {Zucker}, Daniel B. and {Bland-Hawthorn}, Joss and {Hilmi}, Tariq and {S5 Collaboration}},
        title = "{The effect of the deforming dark matter haloes of the Milky Way and the Large Magellanic Cloud on the Orphan-Chenab stream}",
      journal = {\mnras},
         year = 2023,
        month = jan,
       volume = {518},
       number = {1},
        pages = {774-790},
          doi = {10.1093/mnras/stac3108},
archivePrefix = {arXiv},
       eprint = {2205.01688},
 primaryClass = {astro-ph.GA},
       adsurl = {https://ui.adsabs.harvard.edu/abs/2023MNRAS.518..774L}
}

\begin{appendix}

\section{Kinematics}

\begin{figure}[H]
  \centering
  \includegraphics[width=\columnwidth]{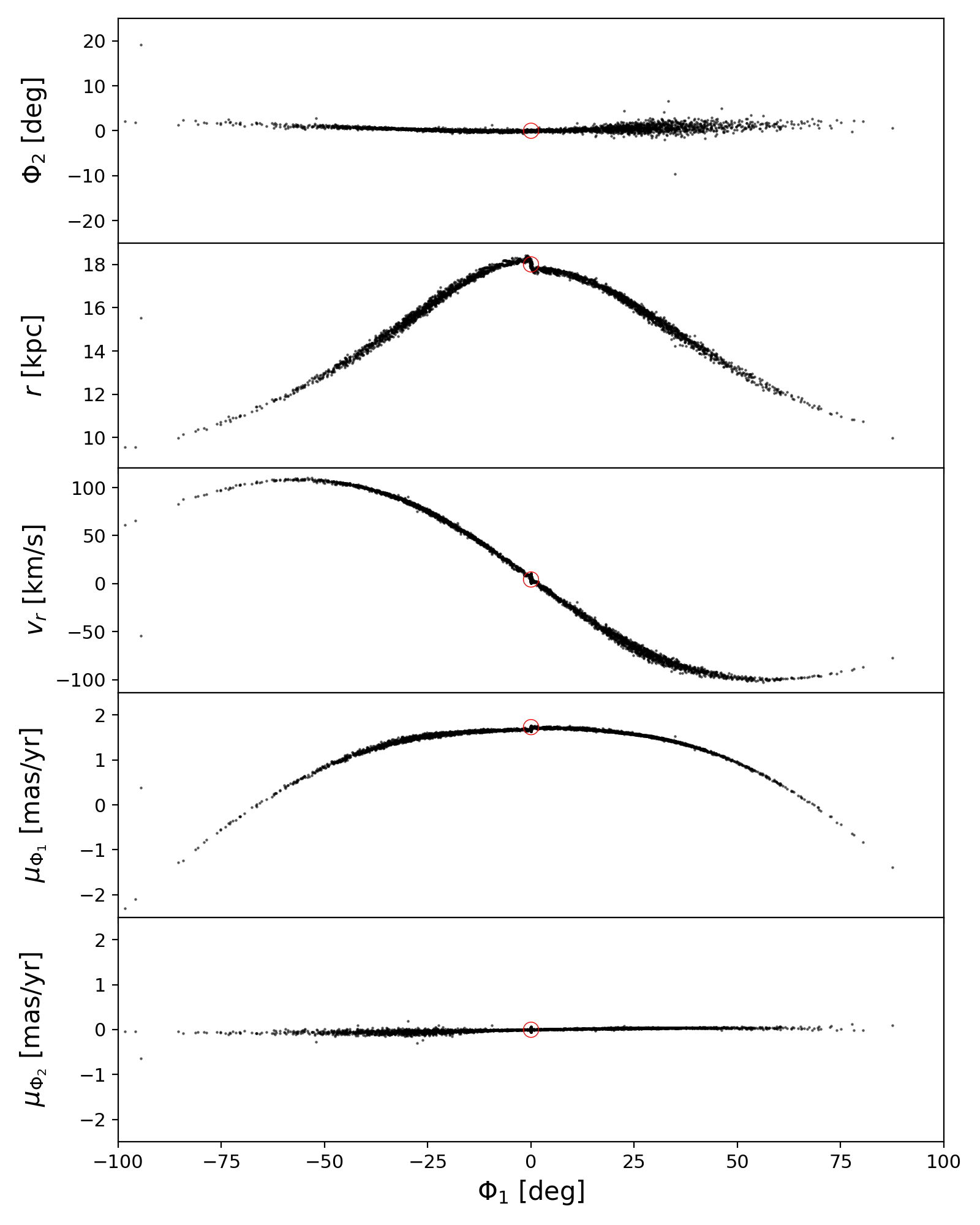}
  \caption{Full set of observables for the same example stream as in Fig.~\ref{fig:example}, in the axisymmetric disc potential. From top to bottom: $\Phi_2$, galactocentric distance $r$, radial velocity $v_r$, and proper motions $\mu_{\Phi_1}$ and $\mu_{\Phi_2}$, all as a function of $\Phi_1$. The progenitor is marked by a red circle in each panel.}
  \label{fig:A1}
\end{figure}

\begin{figure}
  \centering
  \includegraphics[width=\columnwidth]{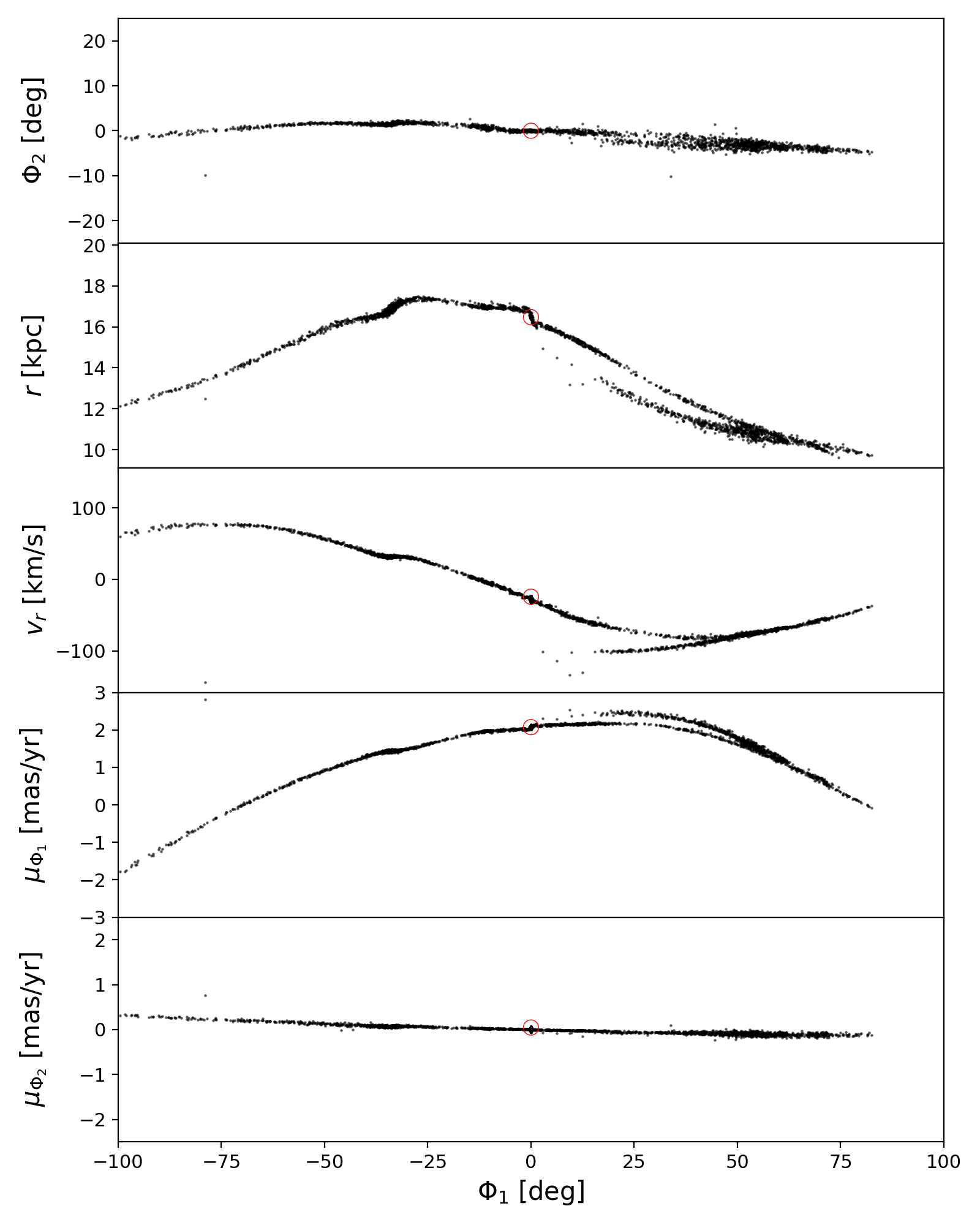}
  \caption{Same as Fig.~\ref{fig:A1}, for the stream evolved with the galactic merger.}
  \label{fig:A3}
\end{figure}

\begin{figure}
  \centering
  \includegraphics[width=\columnwidth]{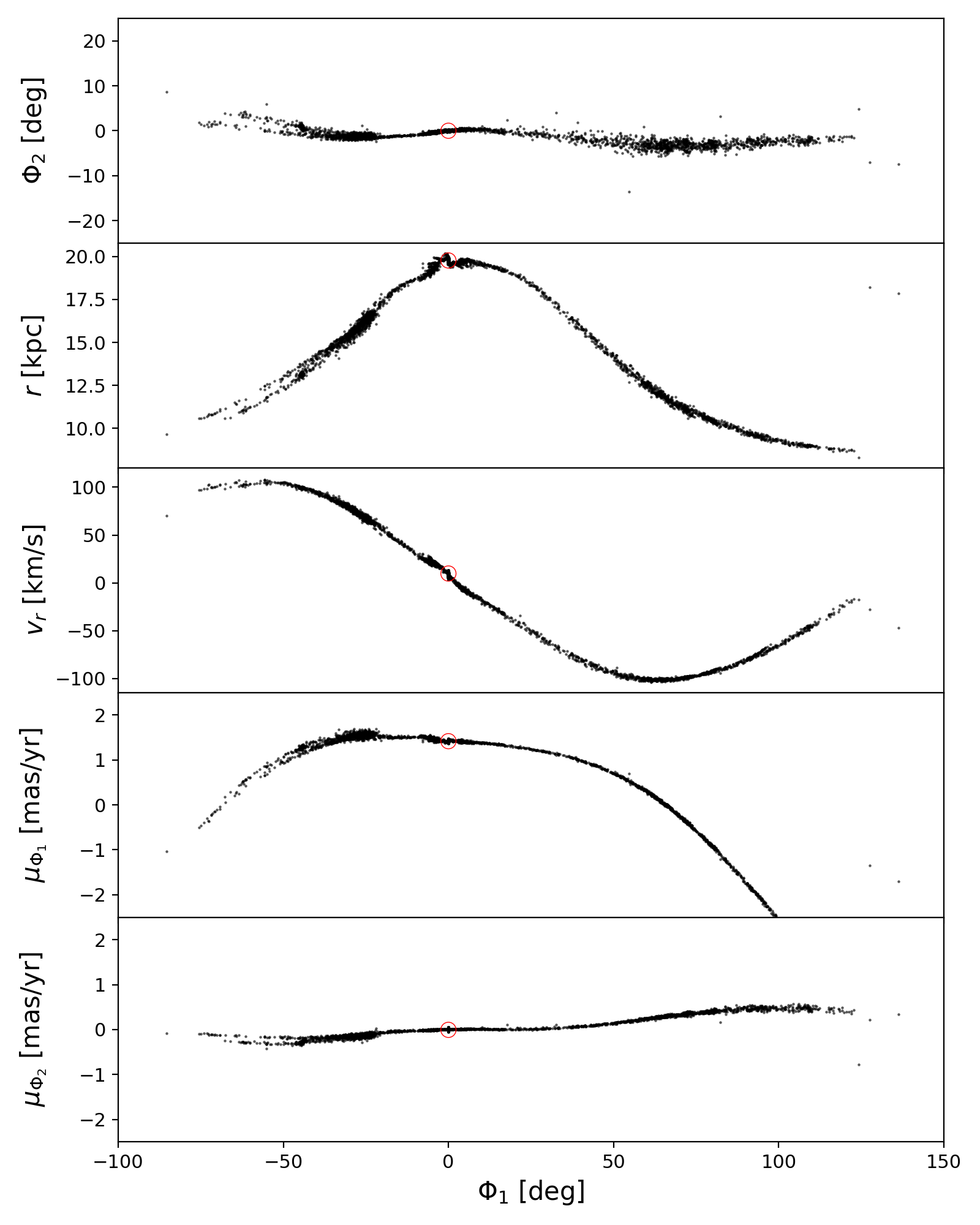}
  \caption{Same as Fig.~\ref{fig:A1}, for the stream evolved with the galactic bar.}
  \label{fig:A2}
\end{figure}

\begin{figure}
  \centering
  \includegraphics[width=\columnwidth]{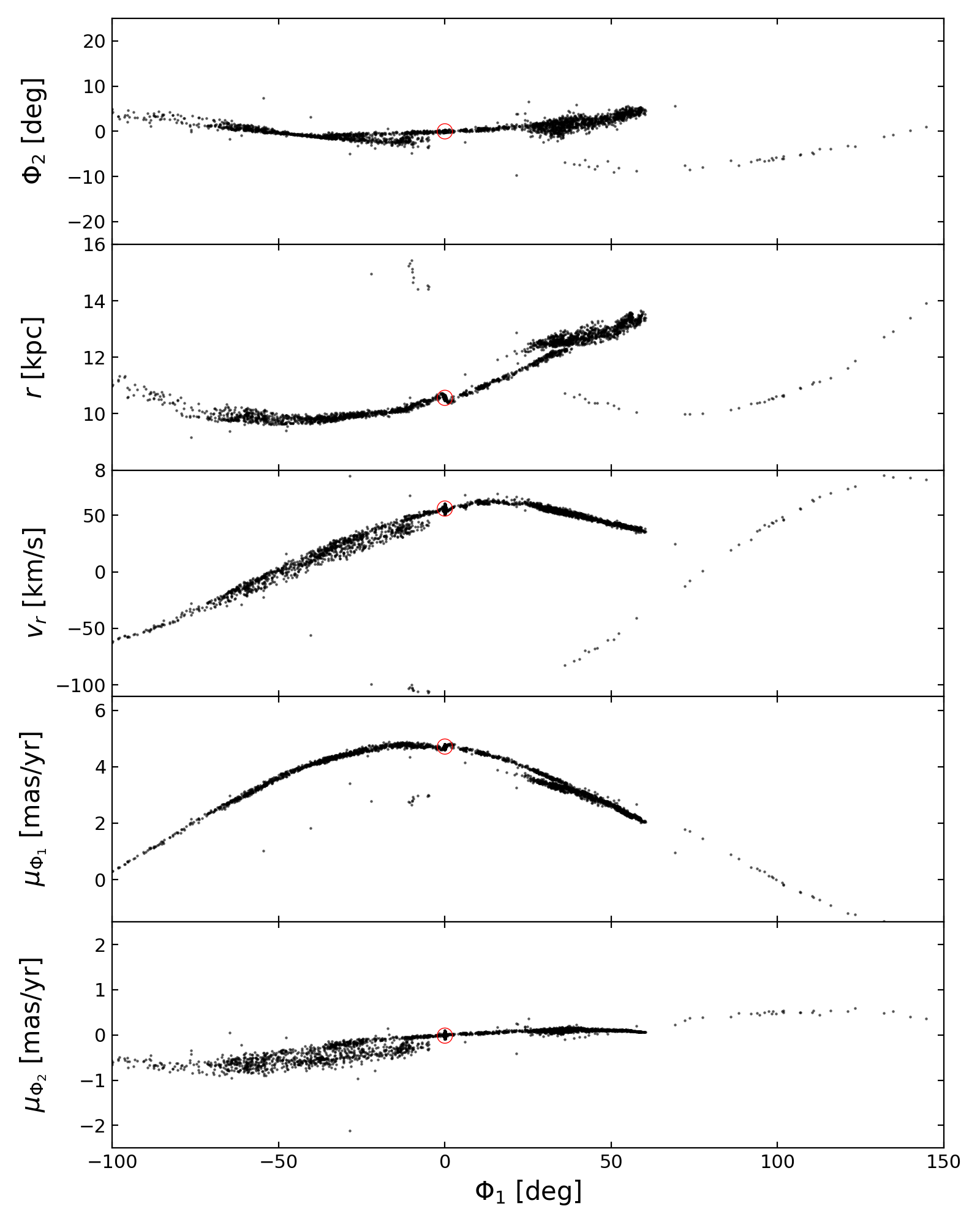}
  \caption{Same as Fig.~\ref{fig:A1}, for the stream evolved with the population of dark matter subhalos.}
  \label{fig:A4}
\end{figure}

Figures~\ref{fig:A1}--\ref{fig:A4} show radial and proper motions for the same stream presented in Figure~\ref{fig:example} for the four configurations: axisymmetric disc, galaxy merger, barred disc, and with dark matter subhalos, respectively. The spur particles overlap with the main stream in proper motion perpendicular to the stream, but are offset in radial velocity and in proper motion along the stream.

\end{appendix}

\end{document}